# Pix2Pix-based Stain-to-Stain Translation: A Solution for Robust Stain Normalization in Histopathology Images Analysis

Pegah Salehi
Image Processing Research Lab
Dept of Computer Eng. & Info. Tech.
RAZI University
Kermanshah, IRAN
Email: pghsalehi@gmail.com

Abdolah Chalechale
Image Processing Research Lab
Dept of Computer Eng. & Info. Tech.
RAZI university
Kermanshah, IRAN
Email: chalechale@razi.ac.ir

*Abstract—* The diagnosis of cancer is mainly performed by visual analysis of the pathologists, through examining the morphology of the tissue slices and the spatial arrangement of the cells. If the microscopic image of a specimen is not stained, it will look colorless and textured. Therefore, chemical staining is required to create contrast and help identify specific tissue components. During tissue preparation due to differences in chemicals, scanners, cutting thicknesses, and laboratory protocols, similar tissues are usually varied significantly in appearance. This diversity in staining, in addition to Interpretive disparity among pathologists more is one of the main challenges in designing robust and flexible systems for automated analysis. To address the staining color variations, several methods for normalizing stain have been proposed. In our proposed method, a Stain-to-Stain Translation (STST) approach is used to stain normalization for Hematoxylin and Eosin (H&E) stained histopathology images, which learns not only the specific color distribution but also the preserves corresponding histopathological pattern. We perform the process of translation based on the "pix2pix" framework, which uses the conditional generator adversarial networks (cGANs). Our approach showed excellent results, both mathematically and experimentally against the state of the art methods. We have made the source code publicly available [1].

## I. INTRODUCTION

Pathological science depends on the examination of microscopic images for the diagnosis of disease based on cellular and regular structures. Most cells are basically transparent, with little intrinsic pigment. Thus tissue Stains are used to confer contrast and reveal the underlying tissue structures and components. In the case of histopathology images, often staining is done by Hematoxylin and Eosin (H&E). hematoxylin is mostly bound to the nuclei (deep purple or blue color), and eosin is mostly bound to the cytoplasm (red color). Several factors can affect the final appearance of the stained tissue, resulting in color variation and intensity in the histopathology images. Some of these sources of change may be due to human skills in sample preparation, protocols between laboratories, tissue fixation stage, and imaging scanners.

Human color perception can easily understand color changes in images, so pathologists can more effectively cope with color variation, but the performance of CAD (Computer-

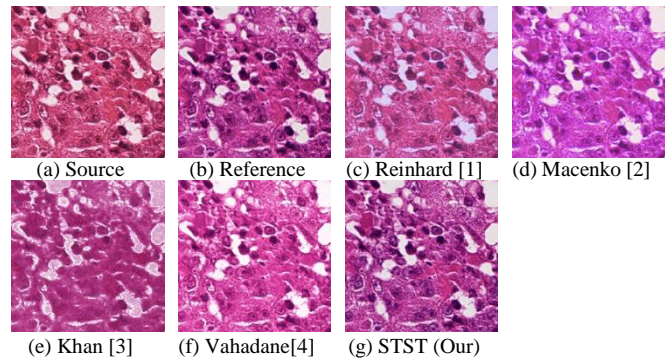

(a) Source    (b) Reference    (c) Reinhard [1]    (d) Macenko [2]

(e) Khan [3]    (f) Vahadane[4]    (g) STST (Our)

Fig. 1. Visual comparison of different stain normalization techniques. The goal of normalization is that the source stain style (a) is looking similar to the Reference stain-style (b).

aided detection) systems decreases dramatically with color change and intensity. Consequently, designing a valid CAD system with the expertise and understanding of pathologists is a challenging task. Hence, the first step of a CAD system is stain normalization, which is a very significant pre-processing step in automated systems. Different stain normalization strategies have been proposed to reduce the inconsistency of stained tissues in automated systems.

Stain normalization should be done in such a way that it maintains good contrast with preserving all the source information in the processed image. Hence a potential drawback of conventional methods is that tissue structures and texture in the original image could be distorted after stain normalization. Also, most of the classic techniques for normalizing all images, use only one user-selected reference image [1]–[6] which has a significant effect on the result of the methods as we show later. For this purpose, we present stain-to-stain translation, a pix2pix-based method, that destroys not only the need for the reference image but also achieves high visual similarity to the ground truth. We train the STST using paired patches of Scanner Hamamatsu and corresponding gray-scale patches. After training, the re-stained patch is compared to the ground truth. Our results show that working with gray-scale images, instead of colored images, generally favor texture-based features, showing then significant improvements in preserving tissue morphology. Furthermore, our proposed method with predict pixels from pixels learn the underlying structures in the tissue. So can say that STST an effective approach as a pre-processing step to reduce the impact that 'non-biological' variations on histopathology data. We compare the results with traditional image processing approaches that have been developed for normalizing histopathology images. The visual appearance obtained from different methods can be seen in Fig.1. It clearly shows that images normalized with STST are very similar to the ground truth.

---

[1] https://github.com/pegahsalehi/Stain-to-Stain-Translation

The rest of the paper is organized as follows. In section II, first Generative Adversarial Networks is introduced, then Previous work done at stain normalization is presented. Moreover, our proposed method is formulated in section III. After that, section IV gives an overview of the image dataset, the STST method implementation details, and the evaluation metrics used for comparing the proposed technique to some methods of stain normalization. Finally, we describe our concluding in section V.

## II. BACKGROUND

In this section, we first introduce Generative Adversarial Networks (GAN), and then briefly describe three major categories of stain normalization algorithms have been proposed in the past.

### A. Generative Adversarial Networks

Generative Adversarial Networks (GANs) [7] are unsupervised generative models that involve two deep neural networks: a generator G and a discriminator D, who are trained simultaneously. It can be considered as a two-player minimax game, where the two players (generator and discriminator) are competing against each other and thus gradually progressing to achieve their goals. The generator is responsible to learn a mapping from a noise vector z in the latent space to output an image in a target domain: $G(z) \to x$, and the discriminator learns to classifies an image as a real image from training image (close to 1) or a fake image produced by the generator (close to 0): $D(x) \to [0.1]$. Both the generator and the discriminator are trained with backpropagation and have their own loss functions. Here, we call them $J_G$ and $J_D$, respectively. The architecture of GANs is illustrated as Figure. 2. During training, the generator learns to produce synthetic samples resembling real images that fool the discriminator, while the discriminator learns to distinguish real and fake samples. To train the networks, the loss function is formulated as following:

$$\mathcal{L}_{GAN}(G.D) = E_{x \sim p_{data}(x)}[log(D(x))] \\ + E_{z \sim p_z(z)}\left[log\left(1 - D(G(z))\right)\right] \quad (1)$$

Where $p_{data}(x)$ denotes the real data probability distribution defined in the data space $x$, and $p_z(z)$ denotes the probability distribution of the latent variable $z$ defined on the latent space $z$, and $E(\cdot)$ represents the expectation.

### B. Previous Work

Previous works published in Stain normalization area often can be broadly categorized into three classes: Stain-separation, template color matching, and style transfer with generative models, which are briefly explained below.

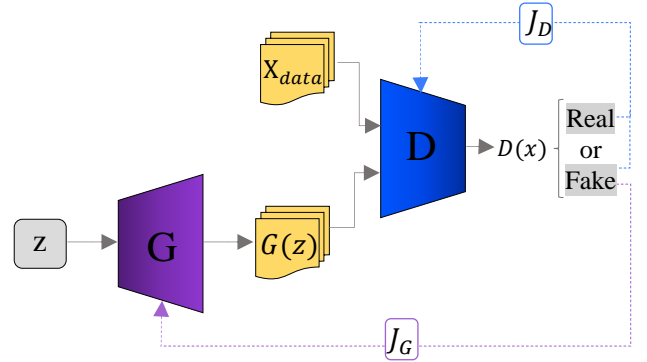

Fig. 2. The architecture of the GAN

*1) Stain-separation:* Since the different stain is possessed with various features in the images, being able to separate information on stains is of central importance. Ruifrok et al. [5] have proposed a novel Stain-separation method in which the stain color appearance matrix was manually estimated by measuring the relative color proportion for R, G, and B channels with only single stained (Hematoxylin or Eosin only) histopathology slide. Such manual estimation of stain vectors limits their applicability in extensive studies. So there are ways to auto-extract colors.

Macenko et al. provide a solution to this problem in [2]. This method assumes that the hematoxylin and eosin stains are linearly separable in the optical density (OD) color space. Hence finds the two largest singular value directions using singular value decomposition (SVD) and projects the OD pixel values onto this plane. However, this kind of method can't always estimate the right stain vectors if strong staining variation is present in histopathology slides [8]. One limitation of this method is the possibility that negative coefficients are obtained in its estimates, which constitutes an invalid biological condition [4]. In another way, Khan et al. [3] using Stain Color Descriptor (SCD) global method obtained overall stain color. Then, to identify the locations where each stain is present, a supervised color classification using the Relevance Vector Machine (RVM) has been applied. However, This is a supervised method in which computation complexity is very much higher. Vahadane et al. [4] developed a stain normalization approach based on sparse non-negative matrix factorization (SNMF) technique to preserve the structural information of the source image. Although the solution space of NMF is reduced by SNMF, its computation complexity is considerably higher. This method also doesn't preserve all color information of the source image.

Although these solutions lead to a better stain estimate, they are limited to image color information, and spatial dependence has been neglected between the structure of the tissue [8].

*2) Template color-matching:* Template Color-matching based algorithms make use of the RGB color spectrum of the image and try to match the channel's levels to that of the reference template. Reinhard et al. [1] proposed to match statistics of color histograms of a reference and source image after transforms of RGB images to the Lab color space. However, when used multiple colors in stained, not exist the assumption of the unimodal distribution of pixels in each channel of Lab color space. Thus, this can result in background areas being mapped as colored regions.

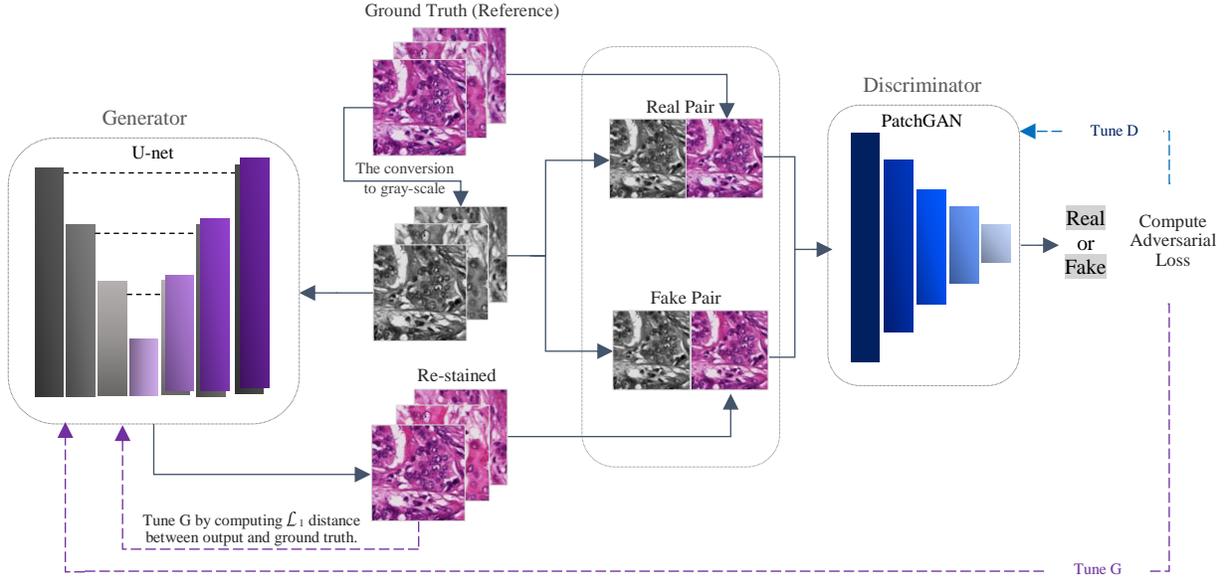

Fig. 3. Illustration of the Pix2Pix framework

3) *Style transfer with generative models:* Generative Adversarial Networks (GANs) [7] (especially cGANs [9]) are a completely different strategy to stain normalization, which was preferred in recent approaches. These generative models handle the problem of stain normalization as a style-transfer problem [10]–[14].

BenTaieb et al. [10] using the concept of style transfer [15], transferred the staining appearance of tissue images across different datasets to avoid color variations caused by batch effects. Nevertheless, in this method, stain normalization does not yield the expected result. In contrast, StainGAN [11], under an unsupervised setting, used CycleGANs [16] to transfer the H&E Stain Appearance between Hamamatsu to Aperio scanners and gained High visual similarity to the target domain. In [13], the photorealism and the structural similarity loss (SSIM) are introduced to keep the structural information unchanged. In [12], first on Input images performed gray-normalization, then GANs are used to transfer a certain style. Also, this technique requires retraining once the reference image changes. In [17] proposed method does not require retraining when the reference image changes but entirely relies on the color of the reference image, not the mean color of the reference image.

## III. THE PROPOSED METHOD

We use Concept image-to-image translation for stain-to-stain translation in Histopathology Images. In the GAN, the generator produces images only from latent variable z. However, in the image-to-image translation task, the generated image must be related to the source image. To solve this, conditional GANs (cGAN) can be employed [9], which takes additional information y as input. For example, a source image is received as additional information for generator and discriminator. The loss function of cGANs is as follows:

$$\mathcal{L}_{cGAN}(G.D) = E_{x \sim p_{data}(x)}[log(D(x.y))] \\ + E_{z \sim p_z(z)}[log(1 - D(G(z.y).y))] \quad (2)$$

Our framework is built using the work of Isola et al. [18] (Pix2Pix), that is an extension of the cGAN. Which learns the mapping from input image to output image along with a loss function to train this mapping. In pix2pix, the $L1$ loss (Eq.3) encourages the generator to produce a sample that resembles the conditioning variable $x$. It is the average value of absolute values of the difference at each pixel between a training image $x$ and the generated image $G(z.y)$.

$$\mathcal{L}_{L1}(G) = E_{x.y.z}[\|x - G(z.y)\|_1] \quad (3)$$

Finally, Eq.3, as an $L1$ normalization term is added to Eq.2, is used as an adversarial loss. The loss function in this work is as follows:

$$\mathcal{L}(G.D) = \mathcal{L}_{cGAN}(G.D) + \lambda \mathcal{L}_{L1}(G) \quad (4)$$

Where λ called lambda, denotes a hyper-parameter that controls the weights of the terms. In our case, it is set to 100. During training, minimized for training a generator and maximized for training a discriminator. In other words, the purpose of training is to find the generator $G^*$ obtained by solving the optimization problem:

$$G^* = arg \min_{G} \max_{D} \mathcal{L}(G.D) \quad (5)$$

The pix2pix method requires image pairs in the training phrase that consist of an original image and the corresponding transformed images, which usually are not easy to obtain in the real world. Among the datasets available for histology, there are no fitting images paired with various stain styles of the same sample. So we utilize from the gray-scale patch and the corresponding RGB patch as pair image.

We have a similar architecture with the Pix2Pix: U-net [19] in generator and PatchGAN [20] in discriminator. In the U-net architecture, the encoder layers and decoder layers are directly connected by "skip connection." Since the skip connection can shuttle the low-level information (which are commonly shared between the input and output images) across the bottleneck of the encoder-decoder net. It effectively improves the performance of stain translation. In convolutional PatchGAN, instead of classifying the whole image together, each image is divided into n×n segments, then it is predicted that each part is real or fake. Finally, by averaging all the answers, the final classification is done. In other words, only the structure at a certain scale of patches is penalized. The Pix2Pix framework in our work is illustrated in Figure 3. The weights of the generator updated via both adversarial loss by the discriminator output and L1 loss by the re-stained image output.

Conditional variants of GAN[7], simultaneously train a conditional generator and a discriminator. The generator is trained to generate images (in our case, H&E re-stained images) conditioned on input images (in our case, the corresponding gray–scale images). The discriminator aims to classify whether the H&E re-stained images are real or fake.

IV. EXPERIMENTAL RESULTS

In this section, we compared the STST with five state-of-the-art histological stain normalization techniques: Reinhard[1], Macenko[2], Khan[3], Vahadane[4].

*A. Dataset*

The public Mitosis-Atypia dataset of images of H&E stained breast tissues released as part of the MITOS-ATYPIA ICPR'14 challenge [21]. The dataset consists of 16 histology slides with three different frames per case scanned with an Aperio Scanscope XT scanner and re-scanned with a Hamamatsu Nanozoomer 2.0-HT scanner. A total of consists of 424 frames at x20 magnification. We cropped each frame x20 of the Hamamatsu scanner into 30 patches of size 256×256 pixels, that finally obtained a total dataset of 12720 non-overlapping patches. For the training set, we extract 3000 random patches from these patches. Also, for quantitative evaluation, we extract 500 patches from 9720 remaining patches (unseen in the training set).

*B. Implementation Details Stain-to-Stain Translation*

Our method does not require a reference image, but for the state of the art methods, we empirically demonstrate our sensitivity to the choice of a reference image. The STST not only learns the mapping from the gray-scale patch to the re-stained patch but also learn a loss function to train this mapping. Since the training of the discriminator is high-speed compared to the generator, therefore the discriminator loss is divided into two to slow down the training process (see Fig. 4). Both generators and discriminators models are trained with the Adam version of stochastic gradient descent with a learning rate 0.0002, and momentum parameters $\beta 1 = 0.5$. Also, both network weights were initialized from a gaussian distribution with a mean 0 and a standard deviation of 0.02. Every experiment is trained for 30 epochs, and the models are updated after each image, In other words, batch size of 1. We used GPU NVIDIA Tesla P100-PCIe-16GB. After training according to loss values, we select one of the best-stored models of the generator. Then using this model, we able to translate any histopathological image to the Hamamatsu scanner.

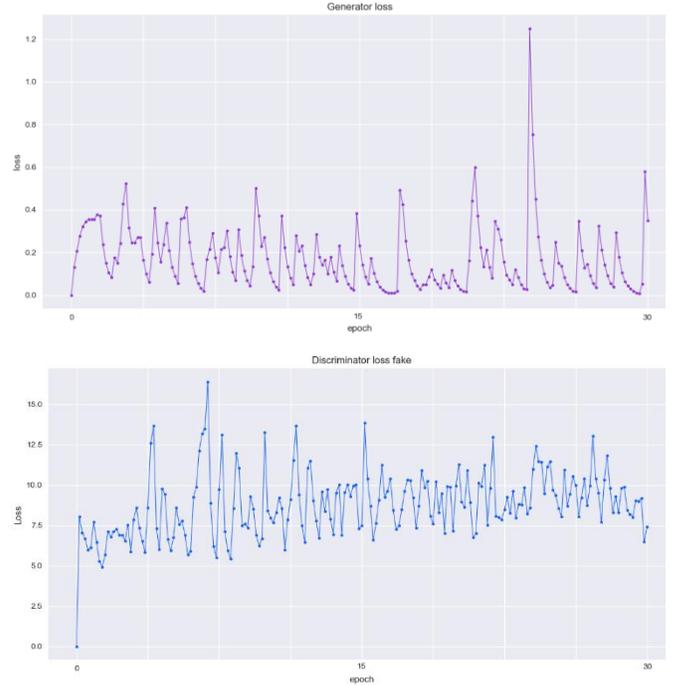

Fig. 4. Discriminator and generator loss during training.

*C. Evaluation Metrics and Results*

Conventional quality metrics (e.g., Full reference metric) are not suitable for histopathology images since usually, the ground truth of histopathology image is not available. Indeed, the ground truth is changed after the normalization process. However, Since we train the STST experiment using paired patches created by Hamamatsu Scanner images and the gray-scale the same images. In this particular case, we can use the Hamamatsu scanner images as the ground truth.

But to have a fair and comprehensive comparison with other ways, the goal should be to be able to normalize the patches from scanner Aperio to Scanner Hamamatsu style, then we compare it with the slides of Scanner H (ground truth). Though the Mitosis-Atypia dataset includes the same tissue sections scanned with both scanners, but because of the difference in the type of scanner, images don't exactly match together. Therefore, part of this value drop in assessing similarity will be due to tiny differences between the two patches. Consequently, to show the excellent result of STST, we examine different evaluation metrics in both the match and non-match ground truth.

TABLE I. DIFFERENT EVALUATION METRICS ARE REPORTED FOR VARIOUS STAIN NORMALIZATION METHODS OF 500 PATCHES ON THE TEST SET (MEAN ± STD.)

| Methods | Unormalization | Reinhard [1] | Macenko [2] | Khan [3] | Vahadane [4] | STST (Our) |
|---|---|---|---|---|---|---|
| SSIM | 0.808 ± 0.034 | 0.807 ± 0.050 | 0.817 ± 0.038 | 0.727 ± 0.063 | 0.841 ± 0.032 | **0.845 ± 0.032** |
| MS-SSIM | 0.908 ± 0.049 | 0.890 ± 0.052 | 0.902 ± 0.052 | 0.794 ± 0.059 | 0.900 ± 0.052 | **0.911 ± 0.051** |
| SCC | 0.260 ± 0.147 | 0.247 ± 0.142 | 0.246 ± 0.141 | 0.131 ± 0.082 | 0.263 ± 0.148 | **0.282 ± 0.157** |
| PCC | 0.887 ± 0.028 | 0.898 ± 0.025 | 0.893 ± 0.031 | 0.849 ± 0.034 | **0.909 ± 0.024** | 0.908 ± 0.027 |
| MSE | 7.65E2 ± 2.28E2 | 11.01E2 ± 4.68E2 | 8.53E2 ± 2.97E2 | 17.66E2 ± 7.38E2 | 11.89E2 ± 3.38E2 | **6.37E2 ± 1.86E2** |
| RMSE | 27.37 ± 4.01 | 32.51 ± 6.69 | 28.81 ± 4.80 | 41.16 ± 8.48 | 34.17 ± 4.71 | **24.95 ± 3.75** |
| PSNR | 19.47 ± 1.26 | 18.06 ± 1.75 | 19.05 ± 1.41 | 16.02 ± 1.77 | 17.53 ± 1.17 | **20.28 ± 1.35** |
| ERGAS | 7.73E3 ± 1.60E3 | 9.08E3 ± 1.89E3 | 8.17E3 ± 2.07E3 | 11.34E3 ± 1.44E3 | 9.80E3 ± 2.21E3 | **6.97E3 ± 1.45E3** |
| RASE | 11.18E2 ± 2.34E2 | 13.12E2 ± 2.75E2 | 11.88E2 ± 3.04E2 | 16.36E2 ± 2.05E2 | 14.27E2 ± 3.31E2 | **10.12 E2 ± 2.12E2** |
| UQI | 0.971 ± 0.012 | 0.946 ± 0.028 | 0.971 ± 0.017 | 0.912 ± 0.032 | 0.960 ± 0.018 | **0.979 ± 0.009** |

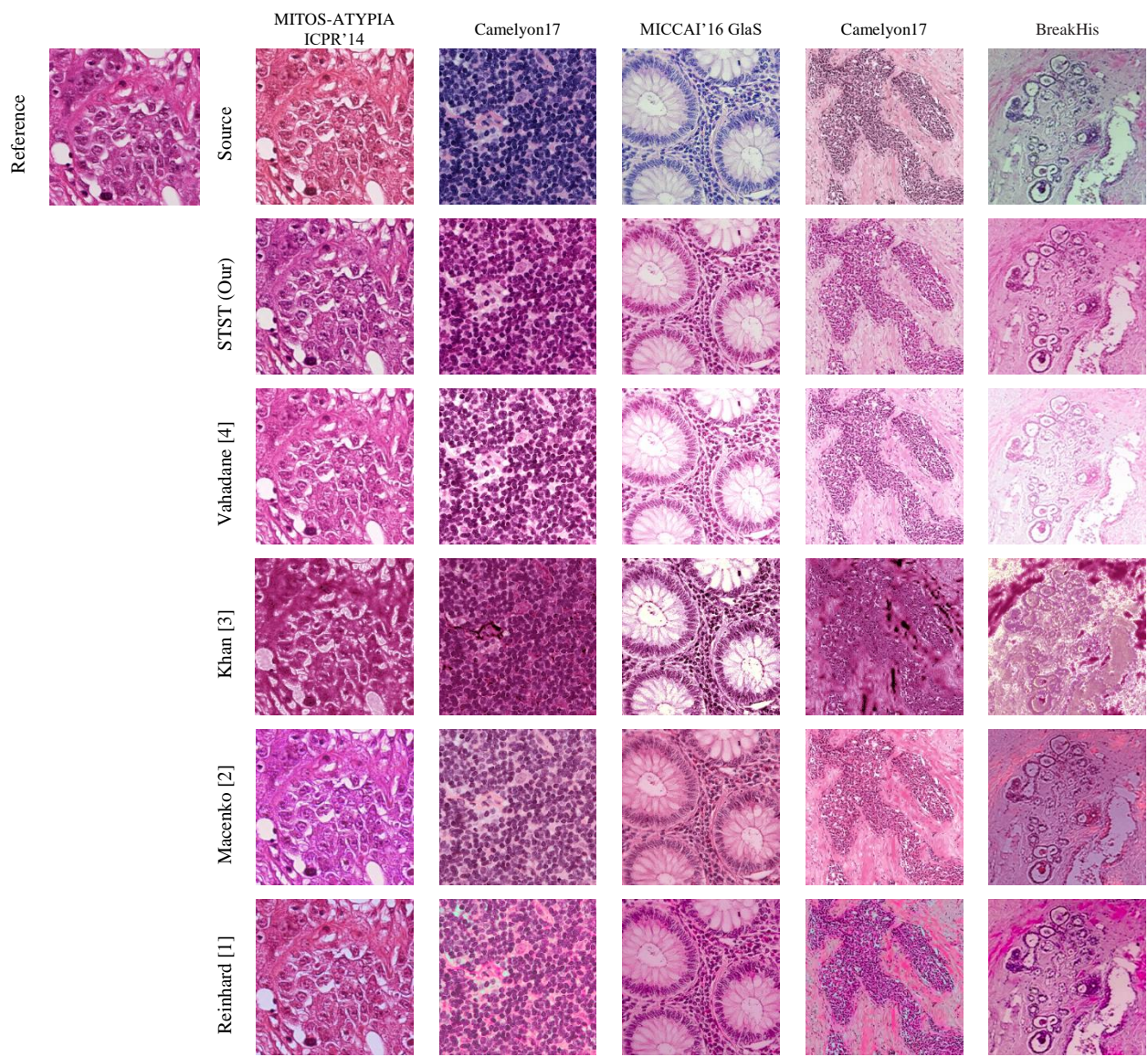

Fig. 5. Comparison of some of the stain normalization methods on H&E stain images from different datasets.

The metrics used for comparison are Structural Similarity index (SSIM) [22], Multi-scale Structural Similarity Index (MS-SSIM) [23], Feature Similarity (FSIM) Index [24], Spatial Correlation Coefficient (SCC), Pearson Correlation Coefficient (PCC) [25], Mean Squared Error (MSE), Root Mean Square Error (RMSE) [26], Peak Signal-to-Noise Ratio (PSNR), Erreur Relative Globale Adimensionnelle de Synthèse (ERGAS), Relative Average Spectral Error (RASE), Universal Quality Index (UQI) [27]. On the other side, to compare the efficacy and better stain separability of the algorithms in extracting the correct stain vectors, the Euclidean distances between the manually determined stain vectors using Ruifrok's method [5] in the ground-truth and the stain vectors computed by different methods.

1) *Patches re-stained are tiny differences with ground truth:* To evaluate the effectiveness of stain normalization algorithms, various metrics based on perceptual similarity and color evaluation have been proposed. However, there still exists an enormous gap between these metrics evaluation and human perception of visual similarity. Hence, we have done both qualitative and quantitative evaluations. From Table I, we can see our method has achieved superior results than all other approaches in all evaluation metrics. Further, in Table II it has shown a better stain separability compared to the ground-truth stain vectors. Also, it has given the processing time in our method is shorter than other methods (Table III).

Most computational metrics are not designed to directly measure the perceptual similarity of the normalized image so that the evaluation results may sometimes be incompatible with the subjective impression. But via visual evaluation, it generally can examine the effectiveness of different methods (Fig. 5).

TABLE II. STAINING SEPARATION COMPARISON (MEAN ± STD.)

| Methods | Staining Separation | | |
|---|---|---|---|
| | $S_H$ | $S_E$ | $S_{Bg}$ |
| Reinhard [1] | 53.02 ± 8.20 | 45.57 ± 15.35 | 46.32 ± 10.65 |
| Macenko [2] | 51.83 ± 8.22 | 29.36 ± 5.13 | 43.16 ± 11.15 |
| Khan [3] | 58.54 ± 9.34 | 46.32 ± 8.91 | 34.84 ± 7.68 |
| Vahadane [4] | 64.78 ± 11.36 | 31.80 ± 6.44 | 34.5 ± 6.31 |
| STST(Our) | **49.03 ± 8.24** | **27.35 ± 5.50** | **34.31 ± 6.23** |

TABLE III. PROCESSING TIME TAKEN TO NORMALIZE THE 500 PATCHES

| Methods | Time (sec) |
|---|---|
| Reinhard [1] | 14.16 |
| Macenko [2] | 73.29 |
| Khan [3] | 1980.14 |
| Vahadane [4] | 553.22 |
| STST(Our) | **9.36** |

2) *Patches re-stained perfectly matched ground truth:* In this section, also both quantitative and qualitative has done assessments. As seen in Fig. 6. show a there is a significant difference in the staining of blood cells compared to cytoplasmic/stromal staining. The STST is able to detect these differences and have done the right staining. Too, the criteria measured in Table IV demonstrate that our method is significantly similar to the ground-truth images and has made more reasonable normalization.

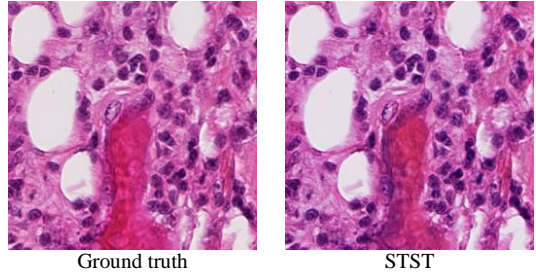

Ground truth          STST

Fig. 6. High capability of the STST in staining blood cells

TABLE IV. COMPARE THE RE-STAINED PATCHES WITH THE MATCHING GROUND TRUTH oF 500 PATCHES ON THE TEST SET (MEAN ± STD.)

| Methods | STST (Our) |
|---|---|
| SSIM | 0.978 ± 0.007 |
| FSSIM | 0.994 ± 0.001 |
| MS-SSIM | 0.998 ± 0.005 |
| SCC | 0.908 ± 0.037 |
| PCC | 0.991 ± 0.002 |
| MSE | 0.74E2 ± 0.27E2 |
| RMSE | 8.51 ± 1.44 |
| PSNR | 29.64 ± 1.39 |
| ERGAS | 2.44E3 ± 0.55E3 |
| RASE | 3.56E2 ± 0.80E2 |
| UQI | 0.996 ± 0.002 |

## V. CONCLUSION

Appearance variation of H&E images can be reduced by adopting proper stain normalization methods that enhance the image contrast. In this paper, inspired by the efficiency of cGANs, that recently has been used as stain normalization methods for histopathological images, we used pix2pix architecture to stain-to-stain translation (STST). Can say that the re-staining process presented in this paper can be viewed as a normalization process where the model learns to re-stained the gray-scale patches with a similar stain-style. Based on evaluation results, we find that STST can provide meaningful colorizations of gray-scale patches and achieved a high perceptually similarity between the ground truth and re-stained image. Moreover, the processing time gained in this method is less than all the methods tested (Table III). So we conclude that effectiveness GANs approaches very outperform the classic stain normalization methods. Hence the STST could potentially be used as a pre-processing step in a histopathologic images analysis pipeline.

ACKNOWLEDGMENT

We want to thank Dr. Babak Ehteshami Bejnordi, for his guidance, time and feedback on this paper.